\documentclass[aps,letterpaper,twocolumn,showpacs,superscriptaddress,nofootinbib]{revtex4-2}

\usepackage[utf8]{inputenc}
\usepackage[english]{babel}
\usepackage[normalem]{ulem}

\usepackage{latexsym}
\usepackage{amssymb}
\usepackage{amsmath}

\usepackage{bm} 

\usepackage{epsfig}
\usepackage{graphicx}
\usepackage{float}
\usepackage{mathtools}

\usepackage{dblfloatfix}

\usepackage{xcolor}

\usepackage{tensor} 

\usepackage[colorlinks=true,allcolors=blue]{hyperref}

\usepackage{comment}

\begin{document}


\title{Gravitational Synchronization in Bosonic Dark Matter Admixed Neutron Stars}

\author{Claudio Lazarte}
\email{claudio.lazarte@uv.es}
\affiliation{Departamento de Astronom\'ia y Astrof\'isica, Universitat de Val\`encia,
Av. Vicent Andr\`es Estell\`es 19, 46100, Burjassot (Val\`encia), Spain}

\author{Nicolas Sanchis-Gual}
\email{nicolas.sanchis@uv.es}
\affiliation{Departamento de Astronom\'ia y Astrof\'isica, Universitat de Val\`encia,
Av. Vicent Andr\`es Estell\`es 19, 46100, Burjassot (Val\`encia), Spain}

\author{José A. Font}
\email{j.antonio.font@uv.es}
\affiliation{Departamento de Astronom\'ia y Astrof\'isica, Universitat de Val\`encia,
Av. Vicent Andr\`es Estell\`es 19, 46100, Burjassot (Val\`encia), Spain}
\affiliation{Observatori Astron\`{o}mic, Universitat de Val\`{e}ncia,
C/ Catedr\'{a}tico Jos\'{e} Beltr\'{a}n 2, 46980, Paterna (Val\`{e}ncia), Spain}




\begin{abstract} 
While the search for dark matter remains a central focus of modern astrophysics and high-energy physics, neutron stars could provide natural laboratories in which the interaction between dark matter and baryonic matter can be studied. In this work we model dark matter as an ultralight bosonic field, which can accrete onto the neutron star and form a composite object bound through gravity. 
Using long-term, numerical relativity simulations in spherical symmetry, we extract and analyze the frequency spectra of the radial oscillation modes of fermion-boson stars. Our simulations reveal that the fermionic and bosonic components synchronize through gravitational coupling regardless of the equation of state, enriching their oscillation spectrum. This synchronization leads to new multi-state scalar configurations and reshapes the hierarchy of the neutron-star radial modes. We further propose a procedure to compute the values of the new dominant modes as a function of the bosonic mass, and discuss how the synchronization mechanism extends beyond spherical symmetry, together with its implications for neutron-star physics and gravitational-wave astronomy.

%
\end{abstract}




\maketitle


\section{Introduction}
\label{sec:intro}
Our understanding of cosmic evolution and large-scale structure relies on dark matter (DM), inferred exclusively from its gravitational effects in galaxy rotation curves, gravitational lensing, and cosmic-microwave-background anisotropies. Although the standard paradigm assumes cold, collisionless particles, no DM candidate has yet been detected. While terrestrial experiments continue to search for DM particles, observations of extreme astrophysical systems provide an important complementary probe \cite{Baryakhtar:2022snowmas}. Among them, neutron stars (NS), with their strong gravity and supranuclear densities, could offer a natural setting in which to search for DM signatures. The past few years have brought forward unique multimessenger insights into NS. Those include precise mass measurements of the most massive millisecond pulsars~\cite{demorest2010two,Antoniadis:2013,cromartie2020relativistic}, LIGO-Virgo-KAGRA gravitational-wave (GW)  detections of compact binary mergers containing at least one NS~\cite{GW170817:letter, GW190425:paper, GW190814:paper, Abac_2024}, X-ray pulse–profile modeling from NICER and XMM-Newton data~\cite{Miller_2019,Riley_2019,Miller_2021}, high-energy observations from H.E.S.S.~\cite{HESS_paper,Doroshenko:2022}, and laboratory constraints from the PREX-I/II neutron skin thickness measurements~\cite{Adhikari:2021_prex}. When interpreted within the standard paradigm of pure NS described by a single (barotropic) equation of state (EoS), several of these constraints remain in tension. This has motivated the consideration of DM–admixed NS, whose modified mass–radius and mass–tidal-deformability relations may simultaneously satisfy multiple multimessenger bounds~\cite{DiGiovanni:2022, Karkevandi:2022, Karkevandi:2024, Sagun:20224z, Shakeri:2024, Pitz:2025, Kumar:2025}. Moreover, this model successfully accounts for the most massive millisecond pulsars observed~\cite{Ivanytskyi:2020,Cipriani:2025}, the tidal deformability extracted from the two confirmed binary NS mergers~\cite{zhang2022,Leung:2022,Diedrichs:2023, Sun:2024,Liu:2024,anh2025algorithmdarkmatteradmixedneutron}, the population of compact objects in the mass gap suggested by recent GW events~\cite{Zhang:2020,Vikiaris:2024_both,Vikiaris:2024asf}, and for the, arguably, lightest NS ever measured~\cite{Sagun_2023}. 

Understanding how DM alters macroscopic properties of a NS may shed light on the amount of DM accumulated inside the star and the microscopic nature of the DM particle, whether fermionic~\cite{Ivanytskyi:2020,Leung:2022,Sagun_2023,Sagun:20224z,Sun:2024,Liu:2024,Kumar:2025,anh2025algorithmdarkmatteradmixedneutron,Vikiaris:2024_both,Vikiaris:2024asf} or bosonic~\cite{Zhang:2020,Karkevandi:2022,zhang2022,Diedrichs:2023,Karkevandi:2024,Shakeri:2024,Pitz:2025,Cipriani:2025}. Further work is however required to identify observables able to distinguish between both situations and break underlying degeneracies. In this paper we take a step in this direction by focusing on the bosonic DM case.

The most common approach assumes that bosonic DM forms a Bose–Einstein condensate~\cite{Guth2015}, with all particles occupying the ground state and negligible thermal excitations. Under a sufficiently strong self-interaction, the condensate becomes effectively isotropic~\cite{Zhang:2020} and can be treated as a perfect fluid described by a DM EoS, allowing equilibrium configurations to be computed through a two-fluid Tolman–Oppenheimer–Volkoff (TOV) system of equations. An alternative description exploits the macroscopic wave-like behavior of the condensate and models the bosonic DM as a complex scalar field. This formulation requires solving a nonlinear eigenvalue problem to obtain equilibrium bosonic configurations. While technically more involved than the fluid approximation~\cite{Diedrichs:2023}, it significantly simplifies dynamical simulations as time integration reduces to solving the Klein–Gordon equation rather than a system of hydrodynamics equations (prone to shocks and discontinuities). Moreover, the scalar-field representation allows to relax the assumption of strong self-interactions and explore a broader class of bosonic families and potentials. When the NS matter is treated as a fermionic perfect fluid and the bosonic component is described by a scalar-field condensate, the resulting object is usually referred to as a fermion–boson star (FBS).

FBS were first introduced in~\cite{Henriques:1989,Henriques:1990a}. Modeling neutrons as a perfect fluid obeying the parametric Chandrasekhar EoS and the bosons as a complex scalar field without self-interactions, \cite{Henriques:1989,Henriques:1990a} found that to achieve comparable energy densities for both types of matter, an ultralight bosonic particle with mass $m_B \sim 10^{-10}\ \mathrm{eV}/c^2$ is required. Subsequent analyses confirmed stable solutions under linear~\cite{Henriques:1990b} and nonlinear perturbations, employing polytropic~\cite{Valdez-Alvarado:2013}, ideal-fluid~\cite{DiGiovanni:2020df,DiGiovanni:2021}, and realistic EoS~\cite{Nyhan:2022}, as well as quartic self-interactions in the bosonic sector~\cite{Valdez-Alvarado:2020,DiGiovanni:2021}. Moreover, \cite{DiGiovanni:2020df} demonstrated a dynamical formation mechanism, where a NS accreting scalar particles cools gravitationally into a FBS with long-lived excited scalar-field configurations with radial nodes. This result motivated a nonlinear stability analysis of first-excited equilibrium configurations~\cite{DiGiovanni:2021}, showing that a sufficient amount of fermionic matter can stabilize otherwise unstable models. Remarkably, \cite{Valdez-Alvarado:2013} found that evolutions of ground-state FBS exhibit new radial modes in the fermionic sector, arising from the gravitational coupling to the bosonic condensate. In addition, \cite{Brito:2015,Brito:2016} studied a DM-admixed NS model with a real scalar field, obtaining quasi-equilibrium configurations in which the matter components oscillate jointly, driven solely by the scalar field. Such a gravitational synchronization remains yet unexplored when the bosonic DM component is modeled as a {\it complex} scalar field.


In this paper we examine the dynamical gravitational synchronization of the fermionic and bosonic sectors in FBS, considering both configurations produced by the accretion of a complex scalar-field cloud onto a NS and equilibrium models under nonlinear perturbations. We find that synchronization reshapes the oscillation spectrum in characteristic ways, suggesting a promising route to constrain bosonic DM using NS asteroseismology. 
%
%


\section{Framework}\label{sec:framework}

Our analysis concerns the evolution of a cold NS admixed with bosonic DM, both only minimally coupled through gravity. The NS is modeled as a fluid, and the bosonic DM as a complex scalar field $\Phi$.  We adopt the action with Lagrangian density, 
\begin{equation}\label{eq:Lagrangian}
    \mathcal{L} = \frac{R}{16\pi}  -\frac{1}{2}\left[\; \nabla_\nu\bar{\Phi}\nabla^\nu\Phi + \mu^2 |\Phi|^2\; \right]\  + \mathcal{L}^{\rm fluid}\;,
\end{equation}
where the bar indicates complex conjugation, $|\Phi|^2=\bar{\Phi}\Phi$, and $\mu=m_Bc/\hbar$ is the mass parameter. Throughout, we use units with $G=c=M_\odot=1$. Variation of this action with respect to $\bar{\Phi}$, yields the Klein-Gordon equation $\nabla_\nu \nabla^\nu \Phi - \mu^2 \Phi = 0$, while variation with respect to the metric $g_{\mu\nu}$ gives the Einstein equations $R_{\mu\nu}-\frac{1}{2}g_{\mu\nu}R = 8\pi T_{\mu\nu}$. Since both matter components interact only gravitationally, the total stress-energy tensor is the sum of two contributions $T_{\mu\nu} = T^{\rm fluid}_{\mu\nu} + T^{\rm boson}_{\mu\nu}$.  From~\eqref{eq:Lagrangian} it follows $T^{\rm boson}_{\mu\nu} = \nabla_{(\mu}\bar{\Phi}\nabla_{\nu)}\Phi  - \frac{1}{2}g_{\mu\nu}\left[ \nabla^\lambda\bar{\Phi}\nabla_\lambda\Phi + \mu^2 |\Phi|^2 \right]$, while for the fermionic component we adopt the perfect-fluid prescription, 
$T^{\rm fluid}_{\mu\nu}  =  [\rho(1+\epsilon)+p] u_\mu u_\nu + p g_{\mu\nu}$, where $u^\mu$ is the 4-velocity of the fluid elements, $p$ the pressure, $\rho$ the rest-mass density, and $\epsilon$ the specific internal energy. 
The Bianchi identities and the Klein-Gordon equation imply the energy-momentum conservation law $\nabla_\mu T^{\mu\nu}_{\rm fluid}=0$, which together with 
baryon-number conservation $\nabla_\mu(\rho u^\mu)=0$, provide the hydrodynamic evolution equations for six fluid degrees of freedom (see~\cite{Montero:2012yr} for the explicit equations). Closure is supplied by the fluid thermodynamical behavior through an EoS, for which we adopt the ideal-gas EoS  $p = (\gamma - 1)\rho \epsilon$, with adiabatic index $\gamma=2$.

We restrict ourselves to spherically symmetric configurations and two types of initial data, a bosonic Gaussian cloud accreting onto a NS (see Appendix~\ref{appendix_A} for details) and FBS equilibrium configurations (cf.~Appendix~\ref{appendix_B}). The integration of the Einstein-Euler-Klein-Gordon system is done with our numerical-relativity code, originally developed in~\cite{Montero:2012yr} and extended by~\cite{Sanchis-Gual:2015lje,Escorihuela:2017} to account for complex scalar fields. This code solves Einstein's equations in the Baumgarte-Shapiro-Shibata-Nakamura formalism in spherical coordinates assuming spherical symmetry and the general relativistic hydrodynamics equations in flux-conservative form. With this framework, we obtain second-order convergent simulations, consistent with the code's integration scheme. Further details on the numerical implementation and convergence tests are given in Appendix~\ref{Appendix_C}. 

\begin{figure}[t!]
\centering
\includegraphics[width=0.49\textwidth]{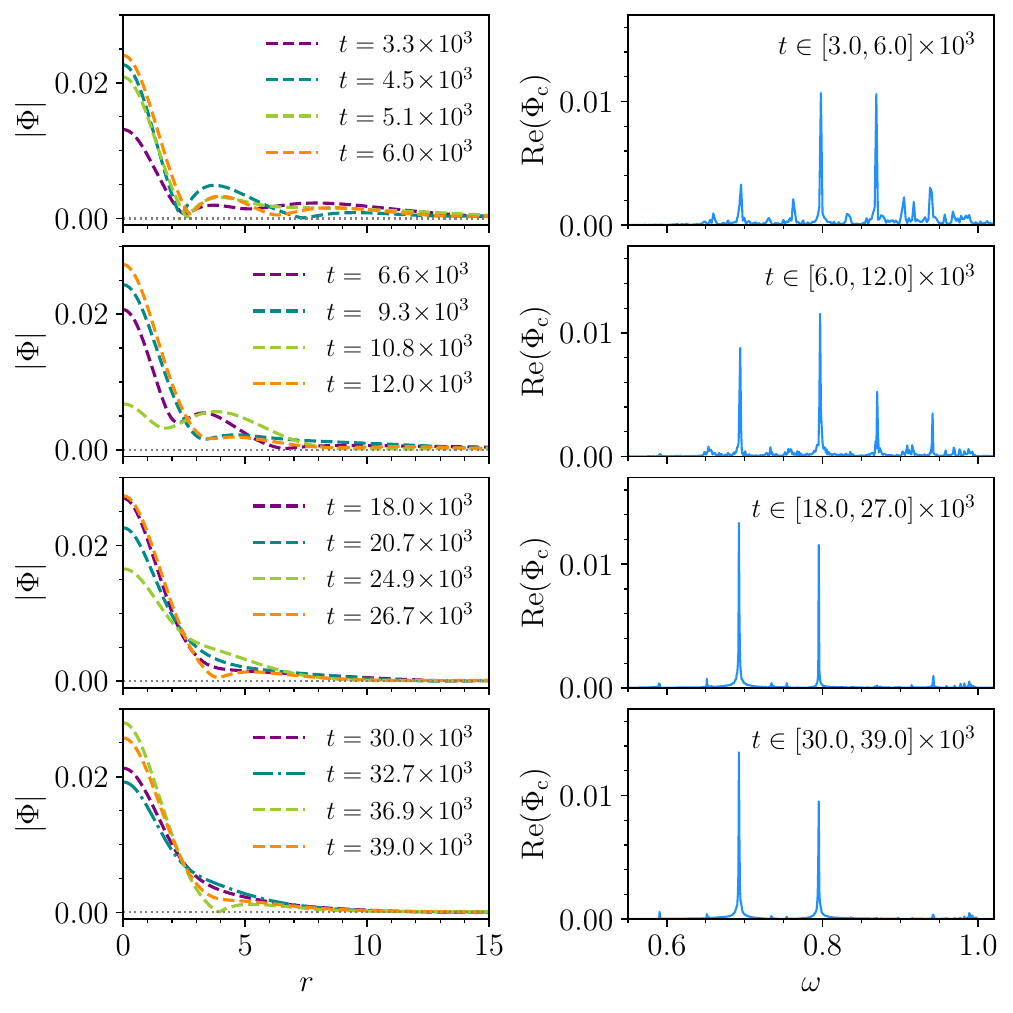}
\vspace{-25pt}  
\caption{Evolution of the radial profiles of $|\Phi|$ (left) and Fourier spectra of the central value of $\rm Re(\Phi)$ (right) for the FBS formation with initial parameters $\rho_c=1.5\times10^{-3}$ and $A_{\Phi}=3.00\times10^{-4}$. Since our code uses geometrized units based on solar masses, physical units are recovered using the conversion factor $1 M_\odot \approx 1.477\;\rm km \approx4.926\!\times\!10^{-6}\rm s.$}
\label{figure:radial_profiles_&_fft_bosonsector}
\end{figure}

\section{Results}\label{resuls}

\subsection{Multi-state bosonic configurations}\label{subsec:multi-state}

We begin with the case $\mu = 1.0\;M^{-1}_\odot$ ($m_B \simeq 1.34\times10^{-10}\, \mathrm{eV}/c^2$) and Gaussian bosonic clouds with initial width $\sigma = 90.0$, initial single frequency $\omega =\mu$, and varying amplitudes $A_{\Phi}$. 
Our simulations extend the evolution time by an order of magnitude beyond those in~\cite{DiGiovanni:2020df,DiGiovanni:2021}. This is necessary to capture the complete dynamical formation of the FBS, ensured once 
 the global quantities of the star reach quasi-equilibrium values: the total mass, computed as the Misner-Sharp mass, and the numbers of fermions $N_{\rm F}$ and bosons $N_{\rm B}$ (see~\cite{DiGiovanni:2020df} for the expressions of the latter two).

A representative evolution of the scalar field after having accreted onto the NS is shown in Fig.~\ref{figure:radial_profiles_&_fft_bosonsector}. Left panels show snapshots of the radial profile of $|\Phi|$ within selected time windows and right panels show the Fast Fourier transform (FFT) of the central value of $\rm Re(\Phi)$ in the same windows. In the latter, the largest peaks from left to right are attributed to the frequencies of the ground (nodeless) state and successive excited states (with one, two, and more nodes). The first row of panels reproduces the time window analyzed in~\cite{DiGiovanni:2020df,DiGiovanni:2021}. Here, we find a two-node excited state dominating the radial dependence. For larger values of $A_\Phi$, one-node or nodeless states dominate in this time window. The next time windows of Fig.~\ref{figure:radial_profiles_&_fft_bosonsector} reveal a gradual transition to the nodeless configuration, reflected in the change in the spectrum peak hierarchy. Although the ground state is initially subdominant, it becomes the main feature in the scalar field spectrum throughout the late-time evolution. However, there is also a persistent first excited state with comparable peak amplitude that contributes with one node in the scalar-field radial profile. This hints to a late-time stable multi-state bosonic configuration. We observe the same multi-state scenario for different values of $A_\Phi$, finding that for larger amplitudes, the mixed stars reach faster the ground-state dominated configurations and the presence of the first-excited state becomes increasingly less relevant. 

\begin{figure}[t!]
\centering
\includegraphics[width=0.49\textwidth]{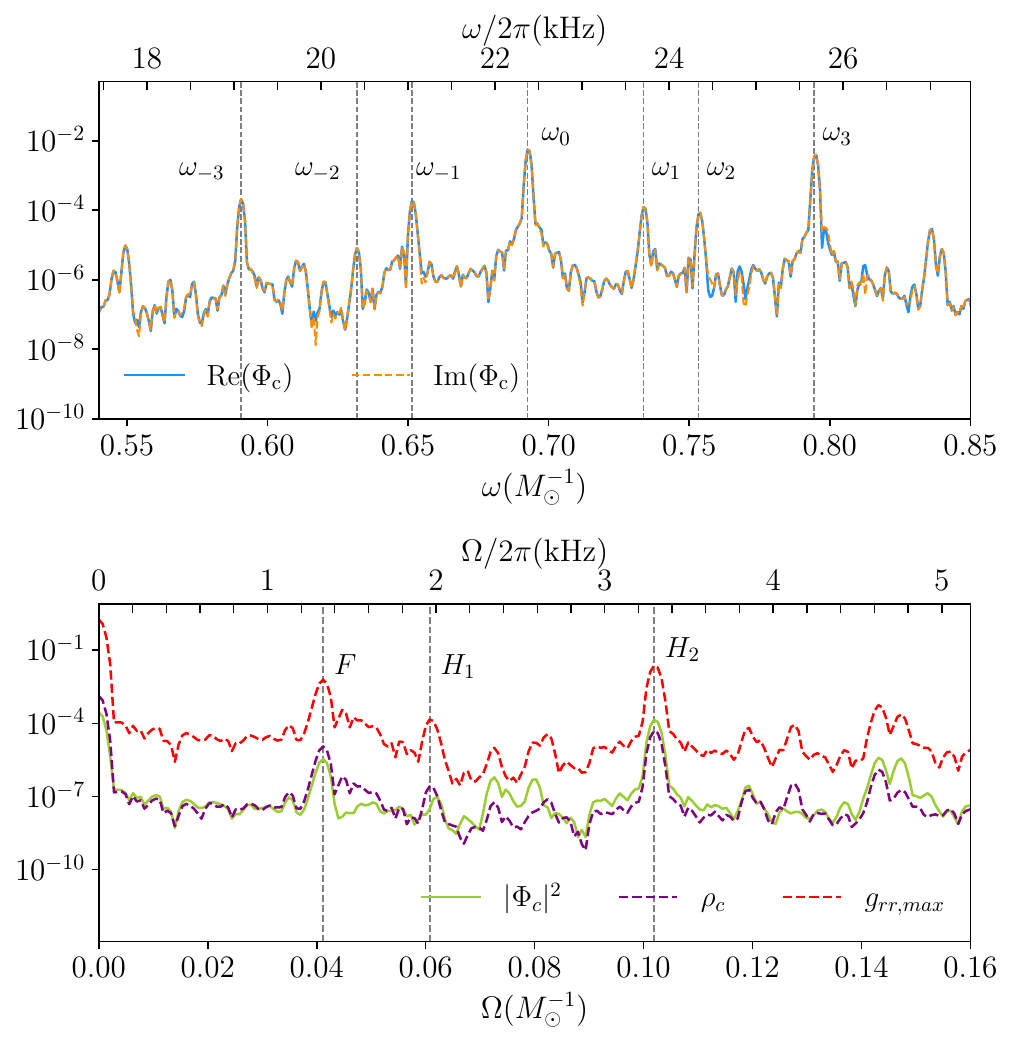}
\vspace{-25pt}      
\caption{FFT of the real and imaginary parts of $\Phi_c$ (top), and spacetime and matter terms that gravitationally synchronize (bottom). The time window used is $[32.7;41.7]\!\times\!10^{3}$ and the corresponding FFT frequency resolution is $\Delta\omega\approx0.7\!\times\!10^{-3}$.} 
\label{figure:MS_FFT_full}
\end{figure}

\subsection{Dynamical synchronization}
An additional Fourier analysis of the real and imaginary parts of the central value of $\Phi$ in logarithmic scale is shown in the top panel of Fig.~\ref{figure:MS_FFT_full}. Here we apply a low-sidelobe window in the data to reduce the spread of spectral-leaked energy from the FFT that can mask weaker peaks placed next to stronger ones.  This procedure unveils several frequency peaks not visible in the linear scale of Fig.~\ref{figure:radial_profiles_&_fft_bosonsector}. The spectrum  shows that the late-time configuration is not only multi-state but it is also characterized by a multi-frequency complex scalar field.
Note that its real and imaginary parts share the same (discrete) different frequencies, with the same radial amplitude for every frequency. This was verified by computing the Fourier spectrum of $\Phi$ at different radial points in the star. 

These findings indicate that the complex scalar field evolves from its initial Gaussian profile into a multi-state, multi-frequency configuration of the form $\Phi = \sum_{n} \phi_n(r) e^{-i\omega_nt}\;, \quad \bar{\Phi} = \sum_{n} \phi_{n}(r) e^{+i\omega_{n}t}$. Then, it follows that $\mathrm{Re}(\Phi)= \sum_{n} \phi_n(r)\cos{(\omega_nt)}, \ \mathrm{Im}(\Phi)= \sum_{n} \phi_n(r)\sin{(\omega_nt)}$ consistently with the behavior we have found. This multi-frequency ansatz implies that 
\begin{equation}\label{exp:multi-freq:module_square}
    |\Phi|^2 = \sum_{n} \phi^2_n + \sum_{n\neq n'} 2\phi_{n'} \;\phi_{n}
    \cos{[(\omega_{n'}-\omega_n)\;t]} \; ,
\end{equation}  
leading to a non-stationary energy density 
 with harmonic radial oscillations. Assuming time-dependence for the whole system, the evolution equations admit solutions for the rest-mass density and metric component $g_{rr}$ as periodic expansions~\cite{Seidel_PhysRevLett.66.1659, Brito:2015},
\begin{eqnarray}\label{eq:fluid_oscillation}
    \rho(t,r) &=&  \rho^{\rm static}(r) + \sum_{n} \rho^{n}(r) \cos{(\Omega_nt)}\; , 
\\
    g_{rr}(t,r) &=& 1 + \sum_n g^{n}_{rr}(r)\cos{ (\Omega_n t)}\; .
\end{eqnarray}
For both types of matter to oscillate radially in sync through their gravitational interaction, the spacetime geometry and both the bosonic and fermionic components must share the same characteristic frequencies. By comparing the cosine phases of~\eqref{exp:multi-freq:module_square} and~\eqref{eq:fluid_oscillation} we derive that
\begin{equation}\label{rel:gravitational_sinchronization}
   \omega_{ n'} = \omega_n \pm \Omega_n \;.
\end{equation}
This is fulfilled to high accuracy, as illustrated in the bottom panel of Fig.~\ref{figure:MS_FFT_full}, which displays the spectra of the module square of the scalar field, the fluid rest-mass density, and the  metric component $g_{rr}$.
Taking into account our frequency labeling, the NS fundamental mode ($F$) and its first two overtones ($H_1, H_2$) map directly onto the scalar-field frequency multiplet, yielding $\omega_{\pm 1}=\omega_0  \pm F$, $\omega_{\pm 2} =\omega_0 \pm H_1$ , $\omega_{\pm 3} = \omega_0\pm H_2$. Although each NS radial mode couples to two scalar-field frequencies, only one frequency corresponds to the first-excited bosonic state (as discussed above). In our case, $\omega_3$ corresponds the first-excited state connecting to the frequency of the second NS overtone $H_2$. Interestingly, in the late-time configuration, this overtone dominates over the fundamental mode of the NS. Synchronization leads to the emergence of $\omega_3$ and of multi-state scalar configurations which, in turn, impacts the oscillation of the NS.

We complement these simulations using different stiffness ideal-gas EoS and cold piecewise polytropic realistic EoS, finding evidence that the gravitational synchronization is unaffected by the thermodynamical treatment of the fluid. Details are reported in Appendix~\ref{appendix_D}. 

\begin{figure}[t!]
\centering
\includegraphics[width=0.49\textwidth]{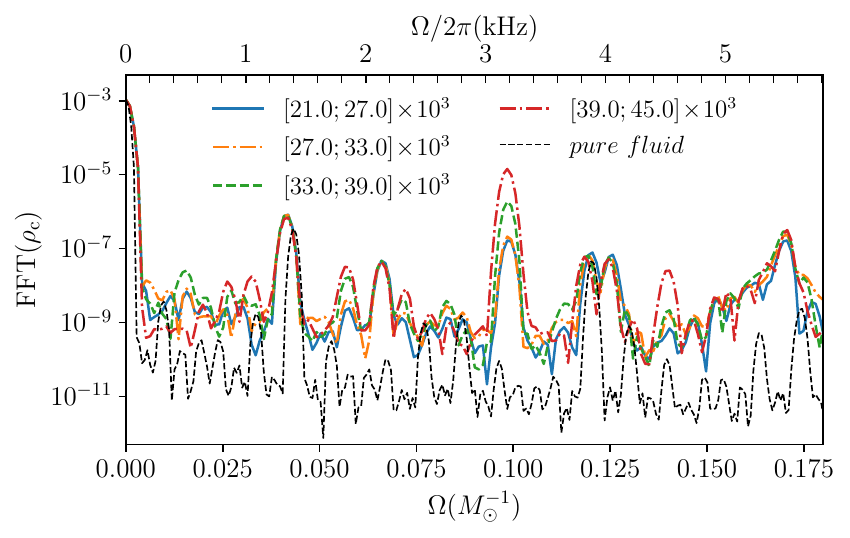}
\vspace{-25pt}       
\caption{NS spectrum showing a resonance in the evolution of an excited equilibrium FBS with one node corresponding to $\rho_c=1.5\times10^{-3}$ and $\phi_{1,c}=1.9\times10^{-2}$.}
\label{figure:ns_spectrum_evolution_1node_trunca}
\end{figure}

\subsection{Gravitational resonance}
 Since the gravitational synchronization we found in the FBS formation is a dynamical phenomenon, where both matter components oscillate with time, we expect it to also occur in evolutions of perturbed equilibrium configurations, in particular in those that initially have a single scalar-field frequency. To check this, we build equilibrium FBS models in both ground state and first excited state configurations, following~\cite{DiGiovanni_2020,DiGiovanni:2021}. These models are evolved up to $t_{\rm final} \sim 4\times10^4$, using the numerical truncation error to perturb them from equilibrium. In particular, the evolution of an excited FBS with one node in the scalar field offers a scenario akin to the dynamical formation case with a dominant first-excited scalar-field state (second row of panels in Fig.~\ref{figure:radial_profiles_&_fft_bosonsector}), yielding a more controlled situation to analyze the synchronization effects on the NS spectra. The results are depicted in Fig.~\ref{figure:ns_spectrum_evolution_1node_trunca}, where a late-time dominance of the second NS overtone is achieved as a resonance effect. In the scalar-field spectrum, this resonance is expressed as the increment of the peak amplitude of the ground-state frequency $\omega_0$. We find that, though this is still subdominant compared to the stable first-excited state with $\omega_1$, both fulfill the synchronization relation~\eqref{rel:gravitational_sinchronization}, $\omega_1-\omega_0 =H_2$, confirming that this evolution undergoes a synchronization process that leads to the appearance of resonances.


\begin{figure}[t]
\centering
\includegraphics[width=0.48\textwidth]{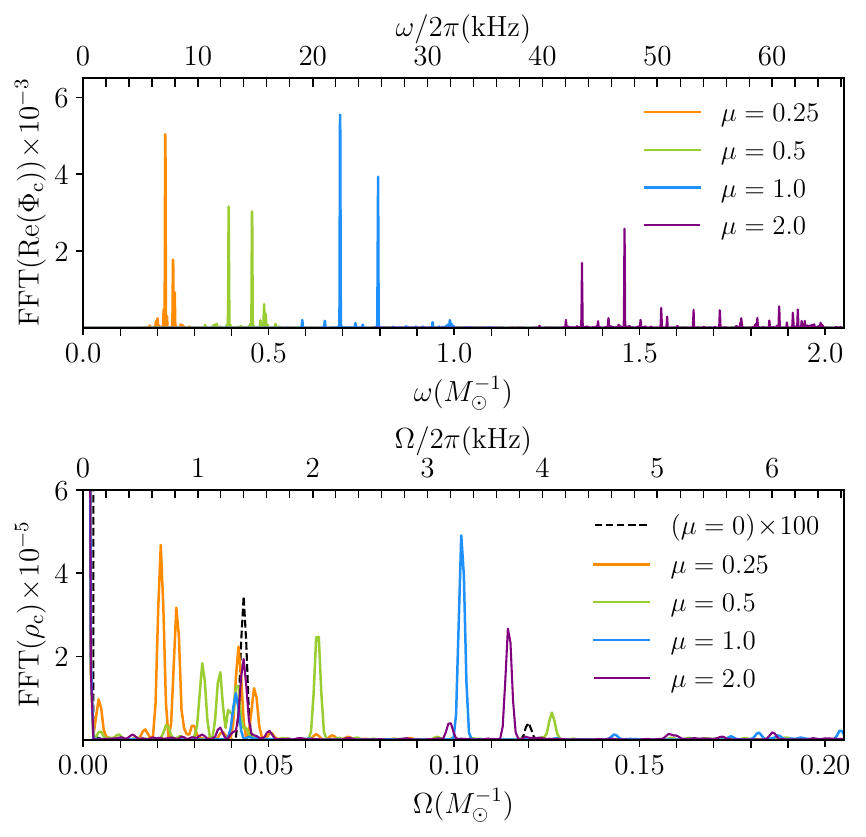}
\vspace{-20pt}  
\caption{Spectrum of $\rm Re(\Phi_c)$ (Top) and the rest-mass fluid density $\rho_c$ (bottom) changing the bosonic particle mass. The time window used was $[33.0;42.0]\!\times\!10^{3}$.} 
\label{figure:different_masses_spectrum}
\end{figure}

\subsection{The role of the boson particle mass}\label{sec:role_of_bosonic_mass}
\begin{figure}[t]
\centering
\includegraphics[width=0.49\textwidth]{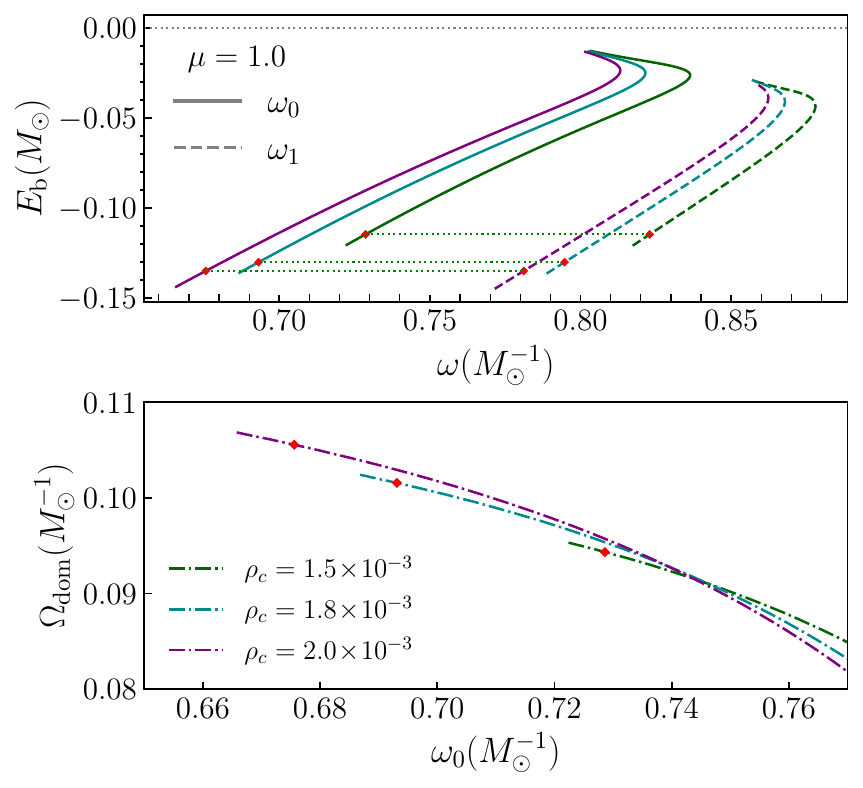}
\vspace{-20pt}  
\caption{Top: Domain of existence of the ground and first-excited state frequencies for $\mu=1$. Solid and dashed lines correspond to ground state and first-excited state frequencies, respectively. The same color denotes to the same $\rho_c$ value in the domains.  Bottom: Resonant frequency obtained as ${\Omega_{\rm dom}}=\omega_0-\omega_1$ from the top panel for different values of $\rho_c$. }
\label{figure:domain_of_existence_resonance}
\end{figure}

The reshaping of the NS radial–oscillation spectrum constitutes an important signature of bosonic DM. It provides a possible approach for detection and conduct inferences of its properties, such as the particle mass. As shown in the bottom panel of Fig.~\ref{figure:different_masses_spectrum}, varying $\mu$ in the FBS formations leads to the appearance of new dominant peaks in their NS spectra. In agreement with our previous dynamical evolutions, these peaks arise from a gravitational resonance between the NS radial modes and the frequency splitting of the coexisting ground and first–excited bosonic states, with characteristic frequencies $\omega_0$ and $\omega_1$, respectively. Thus, with Eq.~\eqref{rel:gravitational_sinchronization}, the associated dominant NS overtone is given by ${\Omega_{\rm dom}} = \omega_0 - \omega_1$. The shift observed in their resonant frequency depends sensitively on the boson mass scale {(See~\cite{Siebel:2001} for the $\mu=0$ case)}. Fig.~\ref{figure:different_masses_spectrum} confirms ${\Omega_{\rm dom}} < \mu$, consistent with the fact that both modes correspond to bound states satisfying $\omega_0, \omega_1 < \mu$. Once $(\omega_0,\omega_1)$ are identified as bound–state frequencies, here extracted dynamically from our formation simulations, they can be mapped onto the domain of existence of single–frequency equilibrium configurations for fixed $(\mu, \rho_c)$, as indicated by the red markers in the top panel of Fig.~\ref{figure:domain_of_existence_resonance}. Remarkably, the two states share the same gravitational binding energy $E_{\rm b}$, enabling us to establish a direct relation between ${\Omega_{\rm dom}}=\omega_0-\omega_1$ and $\omega_0$. We illustrate this relation for $\mu=1$ in the bottom panel of Fig.~\ref{figure:domain_of_existence_resonance}, and we find a similar behavior for $\mu=0.5$. This demonstrates that the procedure is robust across different $\mu$ values and provides a predictive means of identifying the new NS dominant spectral peak produced from gravitational resonance. 

\subsection{Synchronization mechanism beyond spherical symmetry.}
Since we have employed a 1D numerical framework, restricted to the evolution of spherically symmetric solutions, the dynamical gravitational synchronization mechanism manifests itself through a resonance of radial modes in the mixed star. This does not imply, however, that the mechanism is restricted to modes of this kind. The multi-frequency character of a bosonic configuration may also extend to the non-radial spectrum, in which case its energy density, oscillating in synchrony with the fluid density of the mixed star, would obey the same synchronization relations found above. Indeed, let us consider the following two-state complex scalar field, composed of a radial state and a non-axisymmetric state~\cite{Silveira:1995,Schunck1996,Yoshida:1997} with $m=2$ azimuthal dependence,
\begin{equation}\label{eq:non-radial_bosonic_multi-state}
    \Phi(t,\vec{r}) = \phi(r)\; e^{-i\omega t} + \sigma(r,\theta)\; e^{-i(\nu t - 2\varphi)}  \;,
\end{equation}
whose complex conjugate is $\bar{\Phi}(t,\vec{r}) = \phi(r)\; e^{+i\omega t} + \sigma(r,\theta)\; e^{+i(\nu t - 2\varphi)}$. Given these expressions, the modulus squared of the field, $|\Phi|^2 = \Phi \,\bar{\Phi}$, takes the form
\begin{align}\label{exp:multi-freq:module_square}
    |\Phi|^2 & =  \phi(r)^2 + \sigma(r,\theta)^2 \nonumber \\
             &+  2\,\phi(r)\sigma(r,\theta)\cos{[(\nu-\omega)t-2\varphi]}\;,
\end{align}
which introduces non-axisymmetric contributions to the bosonic energy density. For the fluid part, in turn, a rotating NS solution with a developed non-axisymmetric $m=2$ mode has a rest-mass density
\begin{equation}
    \rho (t,\vec{r}) = \rho(r,\theta)  + \varrho(r,\theta)\cos{( \mathcal{W} t-2\varphi)} \;.
\end{equation}
This azimuthal dependence is standard in the context of dynamical fluids. For NS in particular, it describes the harmonic behavior of a fully developed bar-mode instability (see Figs.~7 and~15 of Ref.~\cite{baiotti2007accurate}). Synchronization between the non-axisymmetric $m=2$ modes of the fluid and of the bosonic component then requires their respective cosines to share the same phase. Without loss of generality, we impose this equality at $\varphi=0$, which yields a synchronization relation between the fundamental scalar field frequencies and the modes of the rotating NS that is identical to the one previously obtained for radial modes, but now in the context of non-axisymmetric oscillations,
\begin{equation}\label{exp:sinchro_no_radial}
    \nu =  \omega \pm \mathcal{W}\;.
\end{equation}
This non-axisymmetric mode emits gravitational radiation at the same frequency $\mathcal{W}$ which, as shown for the radial modes, we expect to be amplified by the gravitational resonance between the two components of the mixed star. Should this be realized in nature, the gravitational synchronization mechanism operating in bosonic DM-admixed NS would manifest itself through new frequencies in the gravitational radiation emitted by the star. Preliminary 3D simulations confirm this picture for the evolution of spherical FBS subject to non-axisymmetric perturbations of either the bosonic or the fermionic component. These results reveal a multi-state scalar field component of the form~\eqref{eq:non-radial_bosonic_multi-state}, whose frequencies satisfy the gravitational synchronization relation~\eqref{exp:sinchro_no_radial}, together with the amplification of new gravitational-wave frequencies that are absent in the perturbed pure NS case. A follow-up paper presenting these results is currently in preparation. Following the same logic, we note that rotating FBSs~\cite{Castelo:2024} may trigger additional resonances between rotation-driven NS oscillation modes and the coexisting fundamental frequencies of different axial bosonic states, giving rise to configurations even richer than those derived from Eq.~\eqref{eq:non-radial_bosonic_multi-state}. 

Furthermore, although the bosonic-promoted frequencies of the radial NS spectrum do not radiate GW themselves, they could still be enhanced by nonlinear fluid couplings in realistic, highly dynamical scenarios, thus becoming indirectly observable. We expect this to occur, as it has been shown to for pure NS, through couplings between $m\!=\!2$ and $m\!=\!0$ modes in the differentially rotating hypermassive NS formed in binary mergers~\cite{Stergioulas:2011,Takami:2015,Bauswein:2019}. As a consequence of the synchronization with bosonic DM, a new quasi-radial mode amplified by the gravitational resonance that could dominate the quasi-radial spectrum of the star, might   participate in these couplings  leaving a characteristic imprint on the post-merger GW signal {and its amplifications}~\cite{Passamonti:2006,Passamonti:2007}.

\subsection{New universal relations}\label{sec:universal_relations}
If the robustness of the synchronization mechanism with respect to the fluid EoS reported here for radial modes (see Appendix~\ref{appendix_D}) carries over to non-radial modes, we can follow the same procedure used to construct the bottom panel of Fig.~5: fixing the bosonic mass $\mu$ and choosing a fluid EoS, one finds $\mathcal{W}_{\rm new} = f(\omega_0,\rho_c)$. As is standard in fermion-boson star theory, under these conditions $\omega_0$ depends uniquely on $(\phi_c,\rho_c)$, two parameters of the mixed star which are in turn mapped one-to-one onto $(M_{\rm star},N_{\rm B}/N_{\rm F})$, as shown explicitly in Fig.~1 of~\cite{DiGiovanni:2020df} for $M_{\rm star}$ and in Fig.~2 of~\cite{DiGiovanni:2021} for $N_{\rm B}/N_{\rm F}$. The new NS mode frequency is therefore a function $\mathcal{W}_{\rm new}=f(M_{\rm star},\; N_{\rm B}/N_{\rm F})$ (see Fig.~9 of Ref.~\cite{Valdez-Alvarado:2013} for the same relation for radial NS modes). Once the specific choice of fluid EoS is dropped, an additional parameter is required to fully determine this relation; without loss of generality we choose $\mathcal{W}_{\rm new}=f(M_{\rm star},R_{\rm star},N_{\rm B}/N_{\rm F})$. This last expression could constitute a universal (EoS-insensitive) relation, similar to the one originally proposed in~\cite{Andersson:1996,Nils:1998}, where the frequency of the fundamental (quadrupolar) f-mode $\mathcal{W}_{f}$ and its damping time $\tau_{f}$ are both functions of $(M_{\rm star},R_{\rm star})$ of the pure NS, although in our DM-admixed scenario they may also depend on $N_{\rm B}/N_{\rm F}$. Through future GW asteroseismology we could measure $(\mathcal{W}_{f},\tau_{f},\mathcal{W}_{\rm new})$ and therefore invert the system
\begin{align*}
        \mathcal{W}_{{f}} & =g\;(M_{\rm star}, R_{\rm star},N_{\rm B}/N_{\rm F})\;,\\
        \tau_{{f}} & = h\;(M_{\rm star}, R_{\rm star}, N_{\rm B}/N_{\rm F})\;,\\
        \mathcal{W}_{\rm new} & =f\;(M_{\rm star}, R_{\rm star}, N_{\rm B}/N_{\rm F})\;,
\end{align*}
to obtain the values of $(M_{\rm star},R_{\rm star},N_{\rm B}/N_{\rm F})$ for every $\mu$. In the best-case scenario, either a sufficiently large sample of GW signals from a population of NS admixed with a single species of bosonic DM, or the extraction of a second bosonic-promoted frequency, could break the degeneracy in the bosonic mass. Through $\mu$ and $N_{\rm B}/N_{\rm F}$, this would potentially reveal the presence of bosonic DM fields and, together with $M_{\rm star}-R_{\rm star}$, constrain the properties of nuclear matter for different EoS, as proposed in~\cite{DiGiovanni:2022}. Such new universal relations for bosonic DM-admixed NS might also help to break possible degeneracies with other dark matter models, either fermionic or bosonic, within the two-fluid formalism.


\section{Conclusions}\label{Conclusions}
The extra (complex) scalar degree of freedom in a FBS offers a way to alleviate the tension in NS mass–radius measurements reported in recent multi-messenger observations and nuclear physics experiments (see e.g.~\cite{DiGiovanni:2022} and references therein). This suggests the exciting possibility of detecting ultralight bosonic DM through future compact-star observations. We have explored this idea by analysing both, the dynamical formation (via accretion) of bosonic DM–admixed NS, and evolving equilibrium configurations. Our simulations reveal that a gravitational synchronization between the fermionic and bosonic components of a FBS emerges naturally. This mechanism generates a spectrum of bosonic frequencies that resonate with the NS radial modes, and it enables the formation of long-lived, multi-state, multi-oscillating boson configurations, namely, a DM component with an intricate and complex internal structure inside NS. The multi-oscillating nature of the bosonic components resembles the multi-frequency boson stars proposed in~\cite{Choptuik:2019} for a single-scalar-field scenario, and the stabilization mechanism of multi-field, multi-frequency boson stars in~\cite{Sanchis-Gual:2021}. In addition, our results extend the gravitational synchronization mechanism previously found for mixed stars with {\it real} scalar fields~\cite{Brito:2015,Brito:2016}. Remarkably, we show that synchronization occurs naturally in generic scenarios: it is not imposed in the initial configurations, and, contrary to previous claims, gravitational interaction alone seems to be sufficient to synchronize a complex scalar field with a fermionic fluid. This finding is closely aligned with the paradigm of Kerr black holes with synchronized bosonic hair~\cite{Herdeiro:2014goa,east2017superradiant,herdeiro2017dynamical}. Despite the differences between the two setups, both exemplify dynamical synchronization in astrophysical systems, in which surrounding complex bosonic fields dynamically ``lock'' to the characteristic frequencies of compact objects (neutron stars or rotating black holes), giving rise to long-lived synchronized configurations.

The dominant peaks produced by gravitational resonance in the spectrum of a bosonic DM-admixed NS differ from those of a pure NS. We have proposed an identification of these new radial NS frequencies based on an inspection of the properties of the equilibrium configuration families. This procedure shows that such radial NS frequencies obey a one-parameter relation with the bosonic ground-state frequency, once the bosonic particle mass, the central rest-mass density and the EoS prescription of the stellar fluid are fixed. 

Although our simulations are restricted to spherical symmetry, the mechanism itself is not. We have shown analytically that a two-state bosonic configuration combining a radial and a non-axisymmetric $m=2$ state obeys the same synchronization relation, $\nu = \omega \pm \mathcal{W}$, now linking the fundamental scalar field frequencies to the non-radial modes of the neutron star. Since these modes do radiate gravitational waves, the mechanism would in this case imprint new frequencies directly on the GW signal, a picture supported by preliminary 3D simulations to be reported elsewhere. Even the radial frequencies promoted by the resonance, which do not radiate themselves, could
become indirectly observable through the nonlinear coupling between quasi-radial and quadrupolar modes known to operate in hypermassive remnants of binary neutron star mergers. Should such bosonic-promoted frequencies be measured, we have argued that they may obey universal relations of the form $\mathcal{W}_{\rm new}=f(M_{\rm star},R_{\rm star},N_{\rm B}/N_{\rm F})$ which, combined with the classical f-mode relations, would allow the boson fraction to be inferred alongside the stellar mass and radius, thereby helping to break the degeneracy with other DM-admixed models and to constrain the properties of nuclear matter. Finally, while our study has focused on scalar-field DM, similar synchronization mechanisms may occur for other field-based DM candidates, such as Proca or Dirac fields. If present, the analysis of their distinct gravitational resonances could help to discriminate among models.


\acknowledgments

We thank Alejandro Torres-Forné for his helpful suggestions regarding our implementation of the FFT analysis, Davide Guerra for his help regarding the theory of piecewise-polytropic EoS, and Miguel Alcubierre, Darío Núñez and Carlos Herdeiro for useful discussions. 
N.S.G.\ acknowledges support from the Spanish Ministry of Science, Innovation, and Universities via the Ram\'on y Cajal programme (grant RYC2022-037424-I), funded by MCIN/AEI/10.13039/501100011033 and by ``ESF Investing in your future''. This work is supported by the Spanish Agencia Estatal de Investigación (grant PID2024-159689NB-C21) funded by MICIU/AEI/10.13039/501100011033 and by FEDER/EU, by the Generalitat Valenciana (Prometeo grant CIPROM/2022/49 and Santiago Grisolía grant CIGRIS/2022/164), by the European Horizon Europe staff exchange programme HORIZON-MSCA2021-SE-01 (grant NewFunFiCO-101086251), and by the Center for Research and Development in Mathematics and Applications (CIDMA) through the Portuguese Foundation for Science and Technology (FCT – Fundação para a Ciência e a Tecnologia) under the Multi-Annual Financing Program for R\&D Units,  2024.05617.CERN (\url{https://doi.org/10.54499/2024.05617.CERN}). The authors acknowledge computer resources provided by the Red Espa\~nola de Supercomputaci\'on (Tirant, MareNostrum5, and Storage5) and the technical support from the IT departments of the Universitat de Val\`encia and the Barcelona Supercomputing Center (allocation grants RES-FI-2023-1-0023, RES-FI-2023-2-0002, RES-FI-2024-2-0012, and RES-FI-2024-3-0007).

\bibliographystyle{apsrev4-2}
\bibliography{referencias}



\appendix

\section{Initial data for FBS dynamical formation}\label{appendix_A}

 Following~\cite{Font:2002} the NS initial data corresponds to a TOV static solution 
which provides three radial functions in polar-areal coordinates (two for the metric and one for the fluid), which we transform to isotropic coordinates with line element $ds^2 = -\alpha(r)^2dt^2 + \psi^4(r)(dr^2 + r^2 d\Omega^2 )$, where $\alpha$ is the lapse function and $\psi$ is the conformal factor. We then include a surrounding bosonic Gaussian cloud with initial radial profile 
$\Phi(t=0,r) = A_{\Phi}e^{-r^2/\sigma^2}$ and time derivative $ \partial_t\Phi(t=0,r)=-i\mu\Phi$,  
which guarantees a vanishing scalar-field energy flux and satisfies the time-symmetric initial-data condition. 
To obtain the conformal factor that incorporates the contributions from both matter components, we solve the Hamiltonian constraint $\psi'' + \frac{2\psi'}{r} + 2\pi \psi^5 \mathcal{E} = 0$, with the total energy density in the Eulerian frame given by 
\begin{align}\label{eq:adm-energy-density}
    \mathcal{E} &= \rho(1+\epsilon) +  \frac{1}{2}\left[ \bar{\Pi} \Pi + \frac{\bar{\Psi}\Psi}{\psi^4} + \mu^2\bar{\Phi}\Phi \right] \;  ,
\end{align}
where $\Pi=\alpha^{-1}(\partial_t \Phi-\beta^{r}\partial_r\Phi)$ and $\Psi=\partial_r \Phi$. The initial lapse is the one of the TOV star solution. Our choice of the initial time derivative of the bosonic cloud corresponds to selecting an initial frequency $\omega=\mu$. This frequency ensures that the amplitudes of the real and imaginary parts of the accreting scalar field are equal in the evolution, consistent with the multi-frequency scalar-field ansatz we employ.


 \section{Initial data for equilibrium FBS configurations}\label{appendix_B}
 We construct FBS static models following the procedure implemented in~\cite{DiGiovanni:2020df,DiGiovanni:2021} to which the reader is addressed for further details. Briefly, we employ the same TOV static solution for the NS component as in the dynamical formation case, but adopting a single-frequency scalar-field ansatz $\Phi = \phi_n(r) e^{-i\omega_n t}$, with $n = 0,1,\ldots$ denoting the number of radial nodes. Under this ansatz and the areal-polar slicing condition for the lapse, the Einstein--Euler--Klein--Gordon system reduces to five coupled, first-order radial ordinary differential equations. With suitable asymptotic conditions, this system becomes a nonlinear eigenvalue problem for $\omega_n$. Additional regularity conditions show that the solutions depend on two free parameters: the central rest-mass density and the central scalar-field amplitude, $(\rho_c, \phi_{n,c})$. We solve the resulting problem using a two-parameter, one-dimensional shooting method for $\omega_n$. This procedure yields both the ground-state and first-excited-state configurations, providing the initial data for their evolutions and allowing us to map the domain of existence of the resonant $\omega_1 - \omega_0$ frequency (see Fig.~\ref{figure:domain_of_existence_resonance}).

\begin{figure}[t!]
\centering
\includegraphics[width=0.48\textwidth]{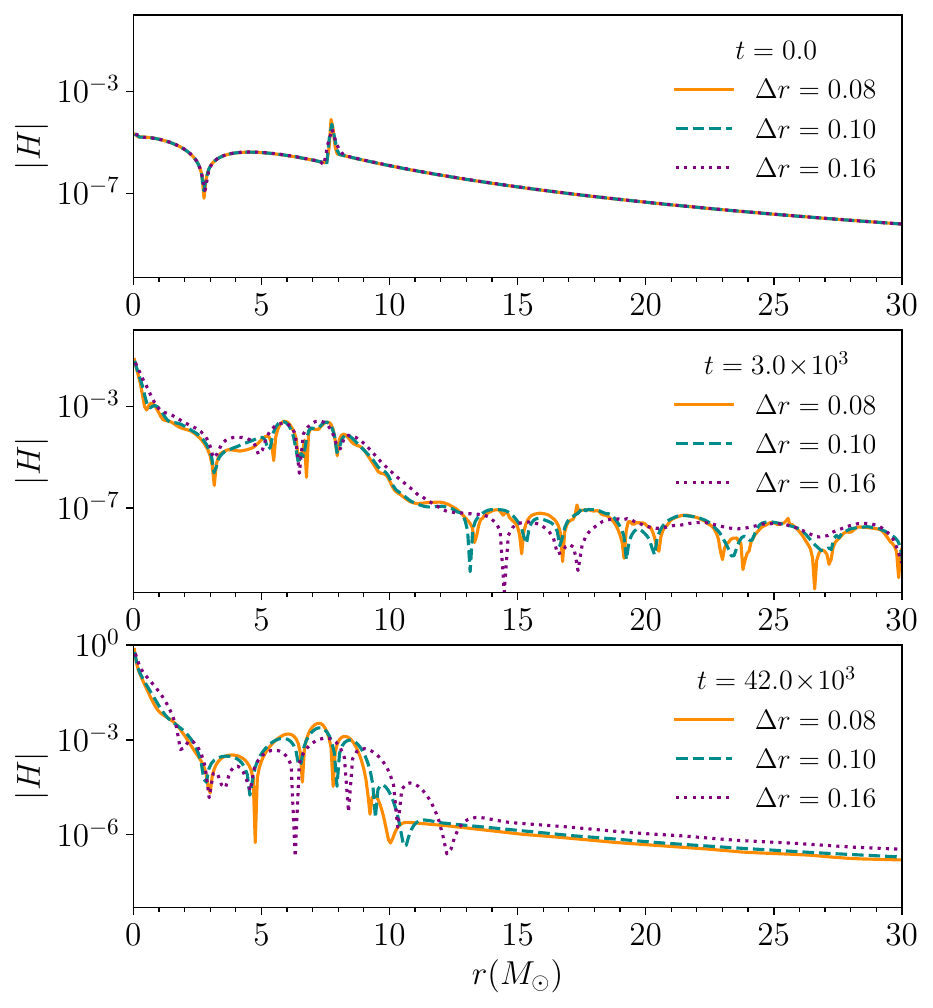}
\vspace{-20pt}  
\caption{Different snapshots of evolution of the Hamiltonian constraint for different resolutions of the fiducial dynamical-formation scenario with $\mu=1.0$. The lower-resolution constraints are rescaled to second-order convergence.}
\label{figure:Hamiltonian1}
\end{figure}

\section{Simulation setup and convergence tests}\label{Appendix_C}
 Our numerical-relativity evolutions employ a radial grid composed of two patches: a uniform mesh extending to an intermediate radius and a hyperbolic–cosine mesh in the exterior. This structure provides the required resolution in the stellar region, centered at the origin, while placing the outer boundary sufficiently far away. The latter is crucial in dynamical-formation scenarios to avoid spurious inward reflections of the scalar field, which could otherwise induce artificial collapse. To suppress such effects, we set the outer boundary at a radius comparable to the total evolution time ($\sim 10^{4}$).

Given the long duration of our simulations, we control computational cost by adopting moderate spatial resolutions. Throughout our simulations we adopt a resolution of $\Delta r = 0.1$ as the default one. To compensate, we employ a Courant–Friedrichs–Lewy factor of $0.15$, ensuring accurate time integration. Time stepping is performed using a partially implicit Runge--Kutta  scheme~\cite{2012arXiv1211.5930C,cordero2014partially}, which effectively handles potential numerical instabilities arising from $1/r$ terms in the evolution equations, without requiring any regularization at the origin. This method yields second-order accuracy, which we observe consistently in all our evolutions.  For the hydrodynamics sector, we use the HLLE approximate Riemann solver together with the second-order MC reconstruction scheme.

\begin{figure}[b!]
\centering
\includegraphics[width=0.48\textwidth]{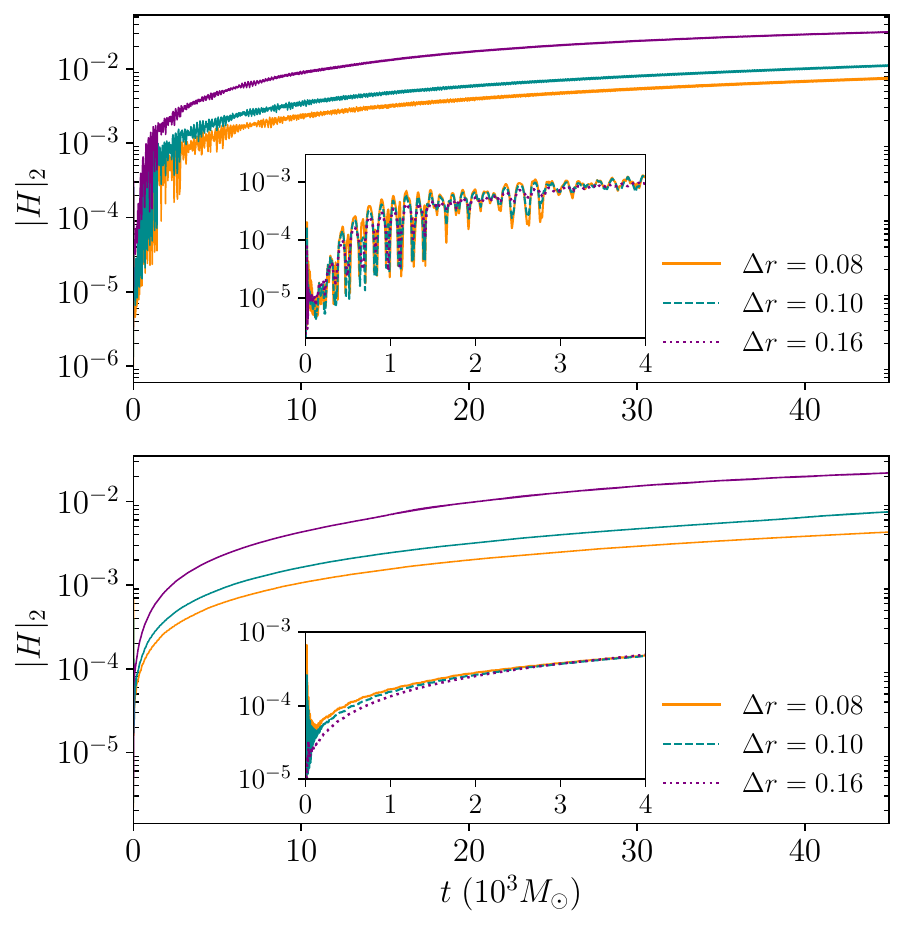}
\vspace{-20pt}  
\caption{Evolution of the $L_2$--norm of the Hamiltonian constraint for the different types of initial data used, dynamical formation (top) and equilibrium FBS configuration (bottom). In the insets, the lower-resolution constraints are rescaled to second-order convergence.}
\label{figure:Hamiltonian2}
\end{figure}

Convergence is explored through the Hamiltonian constraint, given by 
 $   H \coloneq R - {K}_{ij}{K}^{ij} + K^2 - 16 \pi {\mathcal E}$,
where $K_{ij}$ is the (spatial) extrinsic curvature tensor and $K$ its trace.
Fig.~\ref{figure:Hamiltonian1} shows its evolution for the dynamical formation of our fiducial model, demonstrating local second-order convergence. The global convergence test in Fig.~\ref{figure:Hamiltonian2} confirms the convergence behavior for both the dynamical formation model of Fig.~\ref{figure:Hamiltonian1} (top panel) and the equilibrium configuration used in the gravitational-resonance study (bottom panel). The insets show a zoom of the first $4\!\times\!10^3$ time units, where rescaled curves at different resolutions overlap, clearly demonstrating second-order convergence. 

\section{Robustness against variations on the EoS}\label{appendix_D}

In order to probe the robustness of the reported gravitational synchronization mechanism with respect to the choice of EoS, we adopt four additional fluid EoS that differ from the configuration discussed in the main text (ideal-gas EoS with $\gamma=2$). The analysis is divided into two main groups. The first one considers thermal evolutions with the same ideal-gas EoS but different stiffness (different values of $\gamma$). In this case, the initial configurations are obtained using equivalent single-polytropic EoS, where $\gamma=1.9$ and $\gamma=2.1$ correspond to the stiffer and softer EoS according to the mass-radius relations, respectively. The second group is based on cold piecewise-polytropic realistic EoS~\cite{Read2009}, $p=K_i\rho^{\Gamma_i}$ with seven rest-mass density intervals (seven pieces) $\rho_{j-1}\le \rho \le \rho_j$. We consider two benchmark EoS, namely SLy~\cite{Douchin:2001} and DD2~\cite{Typer:2010,Hempel:2010}. Both correspond to softer EoS than our fiducial case.

\begin{figure}[t!]
\centering
\includegraphics[width=0.49\textwidth]{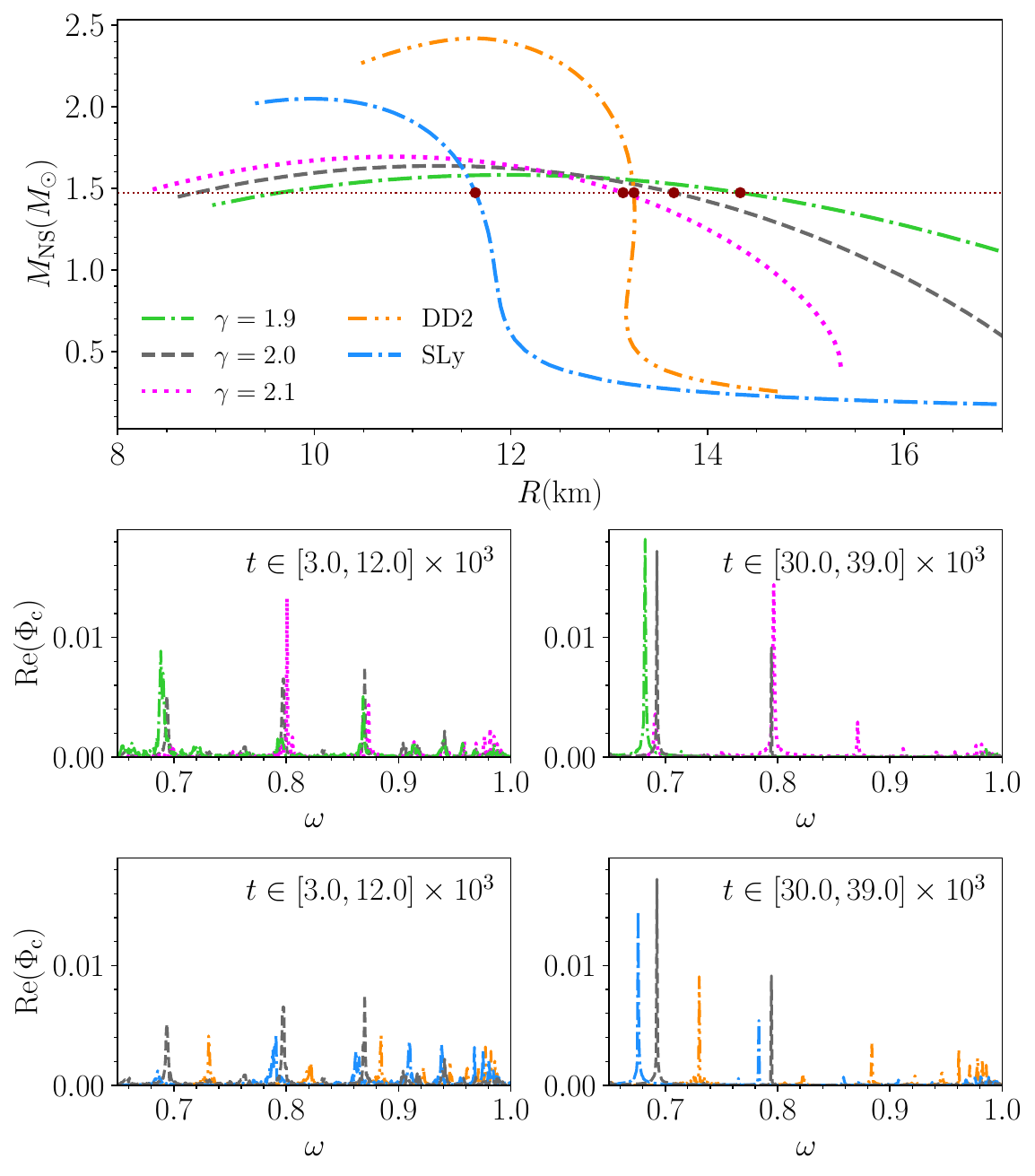}
\vspace{-25pt}  
\caption{Mass-radius TOV relations for all the EoS considered (top).  Panels with initial and late-time windows with bosonic spectra for first EoS group (ideal gas; middle-row) and second EoS group (cold piecewise polytropes; bottom-row). }
\label{figure:Diff_eos_1}
\end{figure}

The top panel of Fig.~\ref{figure:Diff_eos_1} shows the mass-radius relations of TOV stars built with the five EoS considered here. The red points indicate the fiducial NS model used in the FBS formation together with four additional cases with the same gravitational mass. We simulate the bosonic cloud accretion scenario for both types of EoS using the same scalar field initial parameters as in our fiducial case. The middle-row panels of Fig.~\ref{figure:Diff_eos_1} displays two FFT time windows illustrating the evolution of the multi-state scalar configuration obtained for the ideal-gas EoS.  It is observed that the time scale of excited-state dominance over the ground state varies with EoS stiffness; specifically, softer ideal-gas EoS prolong this dominance. For the simulations with realistic EoS, the same multi-state scalar configurations emerges, as shown in the bottom-row panels of Fig.~\ref{figure:Diff_eos_1}, despite a clear change in the frequency peaks is observed in comparison with the evolution of the ideal-gas EoS. These analyses confirm that the appearance of multi-state scalar configurations in the formation of FBS is thus independent of the EoS prescription.

\begin{figure*}[t!]
\centering
\includegraphics[width=0.99\textwidth]{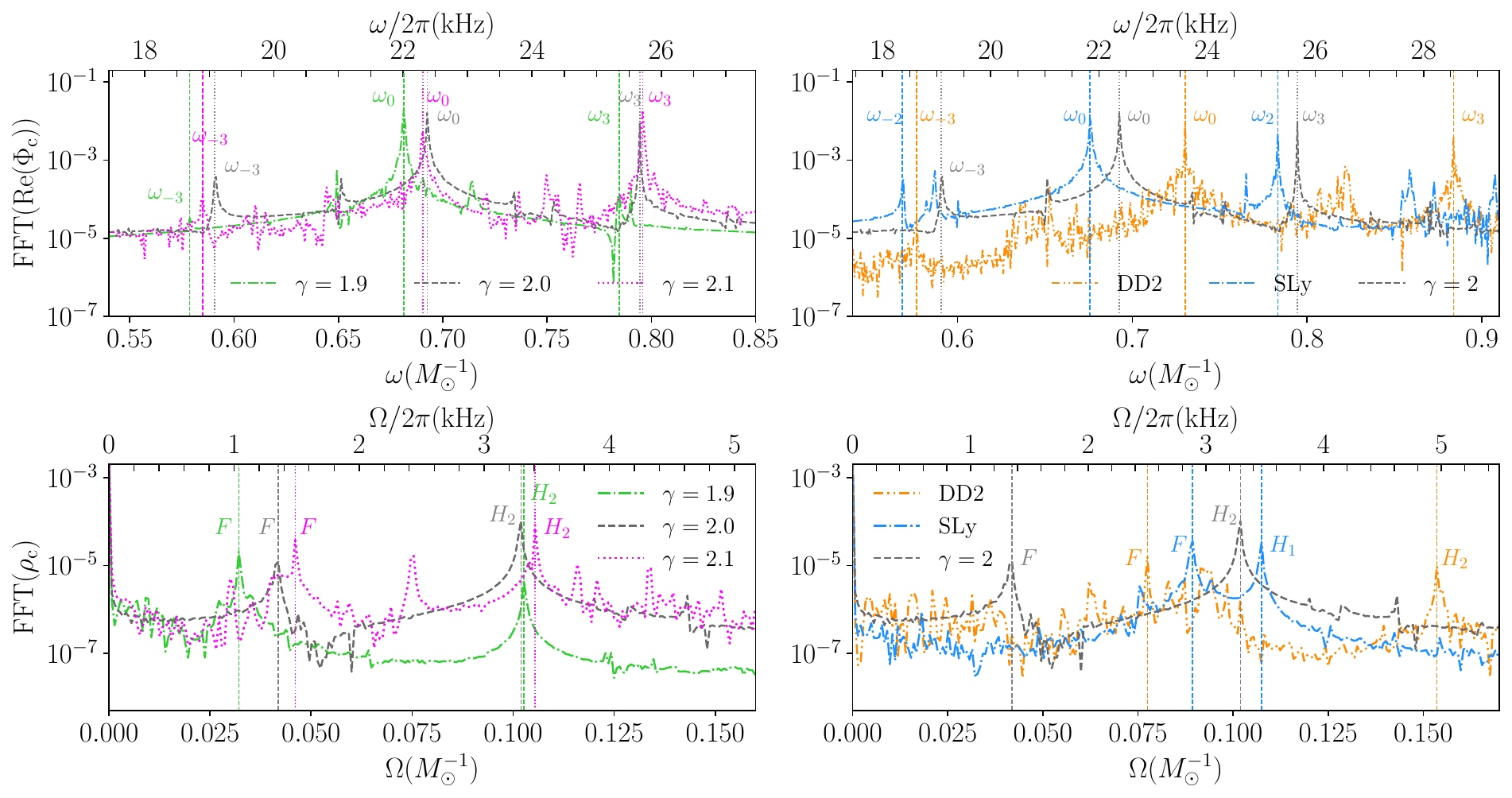}
\vspace{-12pt}  
\caption{Comparison of spectra highlighting the dominant peak frequencies for different EoS: first group (ideal gas; left) and second group (cold piecewise polytropes; right). In all cases, the bosonic-field frequency states (top-row) together with the NS radial modes (bottom-row) satisfy the synchronization relations.}
\label{figure:Diff_eos_2}
\end{figure*}

Finally, we quantitatively test the synchronization relation proposed in Eq.~\eqref{rel:gravitational_sinchronization} using the extracted frequencies of the real part of the multi-state scalar field and the central rest-mass density of the fluid in both EoS groups. The spectra, with the dominant peaks labeled, are shown for each group in the left and right panels of Fig.~\ref{figure:Diff_eos_2}, respectively. We confirm that they satisfy $\omega_{\pm2}=\omega_0\pm H_1$ for the SLy EoS, and $\omega_{\pm3}=\omega_0\pm H_2$ for the DD2 EoS and the two extra ideal-gas EoS. We note that the windowing technique previously applied in the spectra shown in Figs.~\ref{figure:MS_FFT_full} and~\ref{figure:ns_spectrum_evolution_1node_trunca} was not adopted in this figure. By focusing only on the dominant peaks, this technique broadens both the dominant and subdominant peaks, making it difficult to compare different EoS cases. Despite some variations are found, e.g.~the dominance of the second excited state for the DD2 EoS instead of the first one, and despite the expected shifts in NS mode frequency ranges when adopting realistic EoS, all cases unambiguously display the same synchronization between the bosonic and fermionic components, establishing the universality of this gravitational mechanism.

\onecolumngrid
\twocolumngrid


\end{document}